# Electrically Tunable Band Gap in Antiferromagnetic Mott Insulator Sr$_2$IrO$_4$

C. Wang[1,2], H. Seinige[1,2], G. Cao[3], J.-S. Zhou[2], J. B. Goodenough[2], M. Tsoi[1,2]

[1]*Physics Department, University of Texas at Austin, Austin, Texas 78712, USA*

[2]*Texas Material Institute, University of Texas at Austin, Austin, Texas 78712, USA*

[3]*Center for Advanced Materials, Department of Physics and Astronomy, University of Kentucky, Lexington, Kentucky 40506, USA*

**The electronic band gap in conventional semiconductor materials, such as silicon, is fixed by the material's crystal structure and chemical composition. The gap defines the material's transport and optical properties and is of great importance for performance of semiconductor devices like diodes, transistors and lasers. The ability to tune its value would allow enhanced functionality and flexibility of future electronic and optical devices. Recently, an electrically tunable band gap was realized in a 2D material – electrically gated bilayer graphene[1-3]. Here we demonstrate the realization of an electrically tunable band gap in a 3D antiferromagnetic Mott insulator Sr$_2$IrO$_4$. Using nano-scale contacts between a sharpened Cu tip and a single crystal of Sr$_2$IrO$_4$, we apply a variable external electric field up to a few MV/m and demonstrate a continuous reduction in the band gap of Sr$_2$IrO$_4$ by as much as 16%. We further demonstrate the feasibility of reversible resistive switching and electrically tunable anisotropic magnetoresistance, which provide evidence of correlations between electronic transport, magnetic order, and orbital states in this 5d oxide. Our findings suggest a promising path towards band gap engineering in 5d transition-metal oxides that could potentially lead to appealing technical solutions for next-generation electronic devices.**

Tuning material properties electrically is highly desirable for future developments of device physics and associated technologies. Transition metal oxides (TMOs) are promising candidates for such studies[4]. Of particular interest is the iridate Sr$_2$IrO$_4$ (SIO), which is known to have comparable energy scales of spin-orbit coupling, crystal-field splitting, and electron correlations[5][6]. Recently there has been a growing interest in the study of various physical phenomena in SIO, including magnetoelectricity in a canted antiferromagnetic phase[7][8] and non-Ohmic electron transport[9][10]. Nevertheless, little insight has been developed towards a clear understanding of the interconnections between spin-orbit coupling, electron transport, and lattice dynamics in SIO. In particular, transport mechanisms under external electric/magnetic fields in this Mott insulator remain to be addressed.

Here we present a temperature-dependent study of magneto-transport in Sr$_2$IrO$_4$ under dc electric fields up to a few MV/m achieved with the point-contact (PC) technique. Our sample is a single crystal of $Sr_2IrO_4$ (1.5 mm×1 mm×0.5 mm) synthesized via a self-flux technique [11]. The insert to Fig. 1a shows schematically a point contact between a sharpened Cu tip and the crystal. The tip is brought into contact with a (001) surface of the crystal with a standard mechanically controlled point-contact system described elsewhere [12]. The system provides a means to produce point contacts with sizes $a$ (see insert to Fig. 1a) ranging from microns down to a few nanometers. An electrical current is injected through the contact into the crystal and flows (primarily) along the [001] c-axis into a macroscopic Cu electrode on the back side of the crystal. Note that the so-called constriction resistance from a small area near the contact (on the SIO side) dominates the measured resistance over other resistive contributions, including those from the bulk of the crystal and the Cu tip and back electrodes. The latter can be considered as equipotential electrodes because of their relatively high conductivity (Cu vs SIO). When an electric bias is applied between the electrodes, the electrical potential drop occurs essentially in the direct vicinity of the point contact on the SIO side, thus resulting in a locally high electric field $E$ as well as high local current densities $j \sim E$.

Figure 1a shows current-voltage (I-V) characteristics of 10 different point-contacts with zero-bias resistances ranging from 13 kΩ to 27 kΩ measured at T = 77 K. Contact sizes $a$ can be estimated from the measured contact resistance $R$ using a simple model[12] for diffusive transport that gives $R=\rho/2a$, where $\rho$ is the resistivity of SIO. Assuming $\rho \approx 50$ Ωcm at liquid nitrogen temperature[7], this analysis yields $a$ ranging from 4.8 μm – 2.3 μm for $R$ = 13 kΩ – 27 kΩ[13]. The local electric fields and current densities at the highest bias are of the order $E \sim 10^7$ V/m and $j \sim 10^8$ A/m$^2$, respectively. All I-V curves show a similar non-linear behavior: the contact resistance decreases with increasing dc bias as shown in Fig. 1b; the decrease in contact resistance is symmetric at positive and negative biases. From now on we will mostly focus on such resistance vs bias $R(I)$ (as in Fig. 1b) or $R(V)$ plots which highlight even small deviations from a linear I-V behavior. Note that all I-V curves are non-rectifying, which indicates an ohmic contact at the Cu tip/SIO crystal interface.

The observed "S"-shape of the I-V curves is in good agreement with previous SIO studies of bulk crystals[9] and polycrystalline films[10]. Unlike those standard bulk measurements, our point contacts provide a means to probe the electron transport on a local scale in SIO (~point-contact size $a$) subject to extremely high electric fields (~ $10^7$ V/m). In what follows, we examine several established models for explaining the observed non-linear I-V behavior in undoped semiconductors/insulators, including impurities, defects/traps, and interfacial barriers. We will use both current and voltage as fitting parameters since (a) the local current density is expected to scale with the local electric field while (b) the applied voltage is traditionally used to estimate the applied fields in point contacts; but as the applied field spreads over a larger sample space, it may not reflect well on its local nature.

Figures 2a and 2b show $R(I)$ and $R(V)$ data (black), respectively, for a representative point contact ($R$ = 17.5 kΩ; $a \sim 3.5$ μm) together with fits (colored traces) originating from a series of transport models/physical mechanisms (see Part 1 of Supplementary Material for details and further discussion): (i) defect-induced traps in a semiconductor/insulator crystal are often associated with localized electron states within the band gap; the latter alter the crystal's Fermi level and, therefore, its transport properties via so-called space charge limited currents (SCLC), which are expected to lead to an $I \propto V^2$ dependence[14] (dark yellow trace in Fig.2b); (ii) the emission of carriers from traps stimulated by an applied electric field can lead to Poole-Frenkel (PF) currents $j = j_0 e^{-\beta E^{1/2}}$, where $j_0 = \sigma_0 E$ and $E$ is the electric field close to the contact[15]; such currents can become significant at high enough electric fields and lead to non-linear characteristics as shown by green traces in Figs. 2a and 2b, where the applied field is represented by either current or voltage; (iii) a tunneling barrier at the interface between Cu tip and the SIO crystal can promote a decrease in the tunneling junction resistance with increasing bias due to an enhancement of the thermally excited transport across a biased junction (pink curve in Fig. 2b); (iv) a simple Joule heating may lead to a decrease in the crystal's resistivity at an elevated temperature, which can be modeled by the temperature dependence of resistivity $\rho \propto e^{\Delta/2k_B T_{pc}}$ and the bias dependence of the temperature in the contact $T_{pc} \propto V^2$ (or $\propto I^2$)[16] (cyan curves in Figs. 2a and 2b). It is obvious that none of the established physical mechanisms discussed above can provide an adequate agreement with the observed *R(I)* or *R(V)* data over the entire range of applied biases. Below we propose a new mechanism that is consistent with our observations and involves a change of the crystal's band structure under externally applied electric fields via a field-induced lattice distortion.

Since electronic states in 5d transition-metal oxides are extremely sensitive to the overlap of neighboring crystal sites, even a subtle change of the lattice structure may lead to a considerable modification of the crystal's band structure. For instance, lattice distortions induced in SIO by high pressure [17] and epitaxial strain [18] were found to change the effective band gap between 200 and 50 meV at liquid nitrogen temperature. Recent studies of the ferroelectric properties in SIO[19] have found that an applied electric field can induce an electric polarization; the latter may be associated with a field-driven displacement of oxygen anions in Ir-O-Ir bonds. In our experiments, extremely high local electric fields may be sufficient to alter the equilibrium positions of oxygen with respect to iridium ions and induce distortions of the corner shared $IrO_6$ octahedra, thus, provoking modifications of the band structure. We have used an electrically tunable band gap in a simple model to fit our data for the bias-dependent resistance. It was found that the data are well fitted in the entire range of applied bias currents *I* (red curve in Fig. 2a) with the following model:

$$R(X) = A * e^{\frac{\Delta(I)}{2k_B T}}, \Delta(I) = \Delta_0 - B * |I|, \qquad (1)$$

where $\Delta_0$ is the band gap at zero-bias ($I=0$) for a given temperature T, A and B are fitting parameters. Note that when using the applied voltage *V* instead of current *I* in Eq. 1, the fitting shows significant deviations from the observations (red curve in Fig. 2b), which indicates that the current is a better measure of the local electric field in a diffusive point contact. Equation 1 was successfully used to fit the data from contacts with different sizes/resistances (see Part 2 of Supplementary Material). Our model suggests that the band gap in SIO decreases by about 16% at the maximum applied field (at *I* = 3 mA in Fig. 2a). Figure S3 in Supplementary Material illustrates a scenario where an apical displacement of oxygen ions driven by the applied electric field results in an elongation of the Ir-O bonds in the basal plane and promotes a reduction of the band gap. We estimate the field-driven displacement of oxygen ions to be of the order of ~ $2 \times 10^{-3}$ Å, which is ~0.1% of the Ir-O bond length (detailed explanation of this estimation can be found in Part 3 of Supplementary Material). This number is consistent with the previously observed effects of pressure[17] and epitaxial strain[18] on the SIO band gap which give ~10 % reduction in the band gap when the lattice constant changes by ~0.1-1 %.

We further verify our field-effect model (Eq. 1) by measuring the temperature dependence of the I-V curves. Figure 3a shows *R(I)* data (black symbols) measured at temperatures *T* from 83-166 K. The *R(I)* data at different temperatures were fitted by Eq. 1 (red curves). Figures 3b, 3c, and 3d show temperature dependencies of the three fitting parameters A, $\Delta_0$, and B, respectively. The first parameter A is related to the zero-bias resistance of the point contact $R = \rho/2a$, which can be written as $R = 1/2ane\mu$, with carrier density $n = 2(m_e k_B T/2\pi\hbar)^{3/2} e^{-\Delta/2k_B T}$ and mobility $\mu$ of a semiconductor with band gap $\Delta$. That relationship gives A= $(4a * e\mu(m_e k_B T/2\pi\hbar)^{3/2})^{-1}$, which is consistent with the temperature dependence of A in Fig. 3b and yields $\mu \sim 0.1 \, cm^2/Vs$ at 83 K. The values of the second parameter $\Delta_0$ – the band gap at zero-bias – agree well with the ones extracted from the ln(R) vs 1/T data (shown later), and the gap increase with increasing temperature is also consistent with previous results in SIO[7][17][18]. Finally, the third fitting parameter B, which quantifies the magnitude of the field effects, is found to decrease with increasing temperature since at higher *T* the same range of applied currents *I* results in a smaller range of applied electric fields due an increased conductivity of SIO.

Variations of the band gap as a function of temperature and electrical bias can be directly elucidated using the standard temperature-dependent resistivity (contact resistance) measurements at constant biases. Figure 4a shows the zero-bias resistance *R* vs *T* along with ln*(R)* vs *1/T* plot (inset to Fig. 4a) of the same data. The slope of the latter dependence is expected to give the band gap value. Following this standard approach we have extracted the temperature dependence of the zero-bias band gap $\Delta_0$ (Fig. 4b) from the derivative *d(*ln*($R_0$))/d(1/T)* of the data in the inset to Fig. 4a. By performing a similar analysis at different values of the applied bias current, one can extract the bias dependence of the band gap at a fixed temperature. The result of such an analysis at *T* = 167 K shows (Fig. 4c) that the band gap $\Delta$ decreases (from its zero-bias value $\Delta_0$) with increasing bias *I*, in agreement with the proposed field-effect model.

In addition to the continuous variations of the resistance (band gap) as a function of the applied bias, we have observed a reproducible and reversible resistive switching. One can notice small jumps in *I-V* characteristics of different point contacts in Fig. 1. Figure 5a shows an example of such a behavior. Here the black curve shows the *R(I)* sweep from positive to negative biases and the grey curve shows the sweep back. When the applied bias current increases beyond a critical $/I_c/$ value the contact resistance has a step decrease. Following the scenario of the field-effect model, the step decrease may be associated with a field-induced structural transition between two metastable states. The ions being displaced/migrated under high electric fields may encounter potential barriers that need a certain energy to be overcome; the applied electric field modifies the energy landscape and could promote the transition over such a barrier, i.e. switching. We can estimate the variation of the band gap between the two states (below and above $I_c$) from the corresponding decrease in resistance. For instance, the switching shown in Fig. 5a gives a band-gap change of about 0.34 meV. Furthermore, the switching threshold $I_c$ exhibits a clear magnetic field dependence $I_c(\mu_0H)$, which correlates with the point-contact $R(\mu_0H)$ magnetoresistance[13] observed at zero bias (compare Figs. 5b and 5c). The increase in $I_c(\mu_0H)$ with increasing field correlates with the decrease in $R(\mu_0H)$, but their relative changes are quite different and cannot be explained by a field-independent critical voltage $V_c=I_cR$. Previous studies suggested that lattice distortions in SIO may cause a magnetoresistive effect due to a strong spin-orbit coupling[7, 8, 13]. The correlations between the magnetoresistance and resistive switching observed in our present work suggest that the magnetoresistive phenomenon in SIO could originate from the band structure modifications associated with the field-induced lattice distortions.

The interplay between magnetism and electron transport may be further illustrated by studying the effects of bias on SIO magnetoresistance (MR)[13]. The MR was previously[13] associated with a crystalline component of anisotropic magnetoresistance (AMR) and the effects of applied field on the canting of antiferromagnetically coupled moments in SIO. Figure 6a shows a typical MR curve $R(\mu_0H)$ where the applied magnetic field is swept from high positive to high negative fields and back. We have measured such MRs at different values of the applied bias current *I* and found that while the resistance *R* at a constant *H* decreases as a function of *I* (see Fig. 6b) in agreement with the behaviors shown in Fig. 1, the MR ratio, defined as $(R_{max} - R_{min})/R_{max}$ with $R_{max}$ and $R_{min}$ maximum and minimum resistances of $R(\mu_0H)$ sweep, increases with the applied bias. Our observation that MR can be influenced by the applied bias can be taken as another evidence of the idea that magnetic field and electrical bias effects on SIO resistance are entangled and originate from the band structure modifications induced by lattice distortions under magnetic and electric fields. The finding of a higher MR ratio at elevated bias currents may imply that the magnetism-induced lattice distortions can be strengthened by an applied electric field. Finally, the observed MR ratio of about 6-10 % (Fig. 6c) is smaller than the typical resistance variations of 30-60% (Fig. 1b) due to the applied electric field, thus, suggesting that in our point contacts measurements electric fields have a stronger effect on the SIO band structure than applied magnetic fields. A combined effect of electric and magnetic fields on the SIO band gap suggests a promising path towards band gap engineering in 5d transition-metal oxides.

**Acknowledgements**

This work was supported in part by C-SPIN, one of six centers of STARnet, a Semiconductor Research Corporation program, sponsored by MARCO and DARPA and by NSF grants DMR-1207577 and DMR-1122603. The work at University of Kentucky was supported by NSF via grant DMR-1265162.

# References

[1] Oostinga J. B. *et al*. Gate induced insulating state in bilayer graphene devices. *Nature Materials* **7**, 151 (2007).

[2] Castro E. V. *et al*. Biased Bilayer Graphene: Semiconductor with a Gap Tunable by the Electric Field Effect. *Phys. Rev. Lett*. **99**, 216802 (2007).

[3] Zhang Y. *et al*. Direct observation of a widely tunable bandgap in bilayer graphene. *Nature* **459,** 820- 823(2009).

[4] Tokura, Y. & Nagaosa, N. Orbital Physics in Transition-Metal Oxides. *Science* **288**, 462 (2000).

[5] Kim B. J. *et al*. Novel $J_{eff}$ = 1/2 Mott State Induced by Relativistic Spin-Orbit Coupling in $Sr_2IrO_4$. *Phys. Rev. Lett*. **101**, 076402 (2008).

[6] Kim B. J. *et al*. Phase-Sensitive Observation of a Spin-Orbital Mott State in $Sr_2IrO_4$. *Science*, **323**, 1329 (2009).

[7] Ge M. *et al*. Lattice-driven magnetoresistivity and metal-insulator transition in single-layered iridates. *Phys. Rev. B* **84**, 100402 (R) (2011).

[8] Fina I. *et al*. Anisotropic magnetoresistance in an antiferromagnetic semiconductor. *Nature Communications* 5, 4671 (2014).

[9] Cao G. *et al*. Weak ferromagnetism, metal-to-non metal transition, and negative differential resistivity in single-crystal $Sr_2IrO_4$. *Phys. Rev. B* **57**. 11039(R) (1998).

[10] Fisher B. *et al*. Electronic transport under high electric fields in $Sr_2IrO_4$. *J. Appl. Phys.* **101**, 123703 (2007).

[11] Cao G. *et al*. Observation of a Metallic Antiferromagnetic Phase and Metal to Nonmetal Transition in $Ca_3Ru_2O_7$. *Phys. Rev. Lett*. **78**, 1751 (1997).

[12] Jansen, A. G. M., van Gelder, A. P. & Wyder, P. Point-contact spectroscopy in metals. *J. Phys. C* **13**, 6073 (1980).

[13] Wang C. *et al*. Anisotropic Magnetoresistance in Antiferromagnetic $Sr_2IrO_4$. *Phys. Rev. X* **4**, 041034 (2014).

[14] Smith A. Space-Charge-Limited Current in Solids. *Phys. Rev*. **97**, 1538 (1955).

[15] Frenkel J. On Pre-Breakdown Phenomena in Insulators and Electronic Semi-Conductors. *Phys. Rev.* **54**, 647 (1938).

[16] Holm R. *Electric Contacts*, Springer-Verlag, Berlin Heidelberg (2000).

[17] Haskel D. *et al*. Pressure tuning of the spin-orbit coupled ground state in $Sr_2IrO_4$. Phys. Rev. Lett., **109**, 027204 (2012).

[18] Serrao C. R. *et al*. Epitaxy-distorted spin-orbit Mott insulator in $Sr_2IrO_4$ thin films. *Phys. Rev. B* **87**, 085121 (2013).

[19] Chikara S. *et al*. Giant magnetoelectric effect in the $J_{eff}$ = 1/2 Mott Insulator $Sr_2IrO_4$. *Phys. Rev. B* **80**. 140407 (2009).

## Figure Captions

**Figure 1**. Current-voltage characteristics of $Sr_2IrO_4$ crystal in point-contact measurements at $T = 77$ K. (a) Voltage (V) vs current (I) characteristics of 11 point contacts with zero-bias resistances ranging from 13 k$\Omega$ to 27 k$\Omega$ are shown in different colors. (b) Resistance (R) vs current (I) plots for the same 11 point contacts. The insert shows schematically a point contact between a sharpened Cu tip (top light grey) and the crystal (dark grey) on a Cu back electrode (bottom light grey).

**Figure 2** Analysis of the bias dependence of point-contact resistance. (a) Measured point-contact resistance $R$ (black) as a function of applied bias current $I$ together with fits by different models: Joule heating (dashed cyan), Poole-Frenkel (dashed green), field-effect (red). (b) Measured resistance $R$ (black) is shown as a function of applied bias voltage $V$ together with fits by: space charge limited currents (dashed dark yellow), Joule heating (dashed cyan), Poole-Frenkel (dashed green), tunneling barrier (dashed pink), and field-effect model (red).

**Figure 3** Temperature dependence of the point-contact $R(I)$ characteristics. (a) Measured $R(I)$ curves (black) at temperatures from 83 to 166 K together with the field-effect model fits (red). The temperature dependences of A, $\Delta_0$, and B fitting parameters are shown in panels (b), (c), and (d), respectively.

**Figure 4** Band gap characterization by temperature-dependent resistivity measurements. (a) Experimental data of $R$ vs $T$ (open triangles) with an exponential fit (black curve). The insert shows the same data as $\ln(R)$ vs $1/T$ (straight lines are guides for eye). (b) Temperature dependence of the band gap as extracted from the slope of $\ln(R)$ vs $1/T$ in (a). (c) The bias dependence of the band gap at $T=167$ K extracted from the $\ln(R)$ vs $1/T$ dependencies at different bias currents (not shown).

**Figure 5** Resistive switching in point-contact $R(I)$ characteristics. (a) A typical $R(I)$ curve with up- (grey) and down- (black) sweeps of $I$. The switching current is indicated by $I_c$ and arrow. (b) The magnitude of critical current $I_c$ vs applied magnetic field is shown with open circles (solid dots) for up- (down-) sweep of magnetic field. (c) Variations in the zero-bias resistance (magnetoresistance ratio) as a function of the applied magnetic field. All measurements were done at $T = 77$ K.

**Figure 6** Bias dependence of magnetoresistance (MR). (a) A typical MR measurement under low bias current (0.02 mA) at $T =77$ K. (b) Bias dependence of the point-contact resistance at zero applied magnetic field. (c) Bias dependence of the MR ratio found from MR data like in (a) at different biases.

**Figure 1**

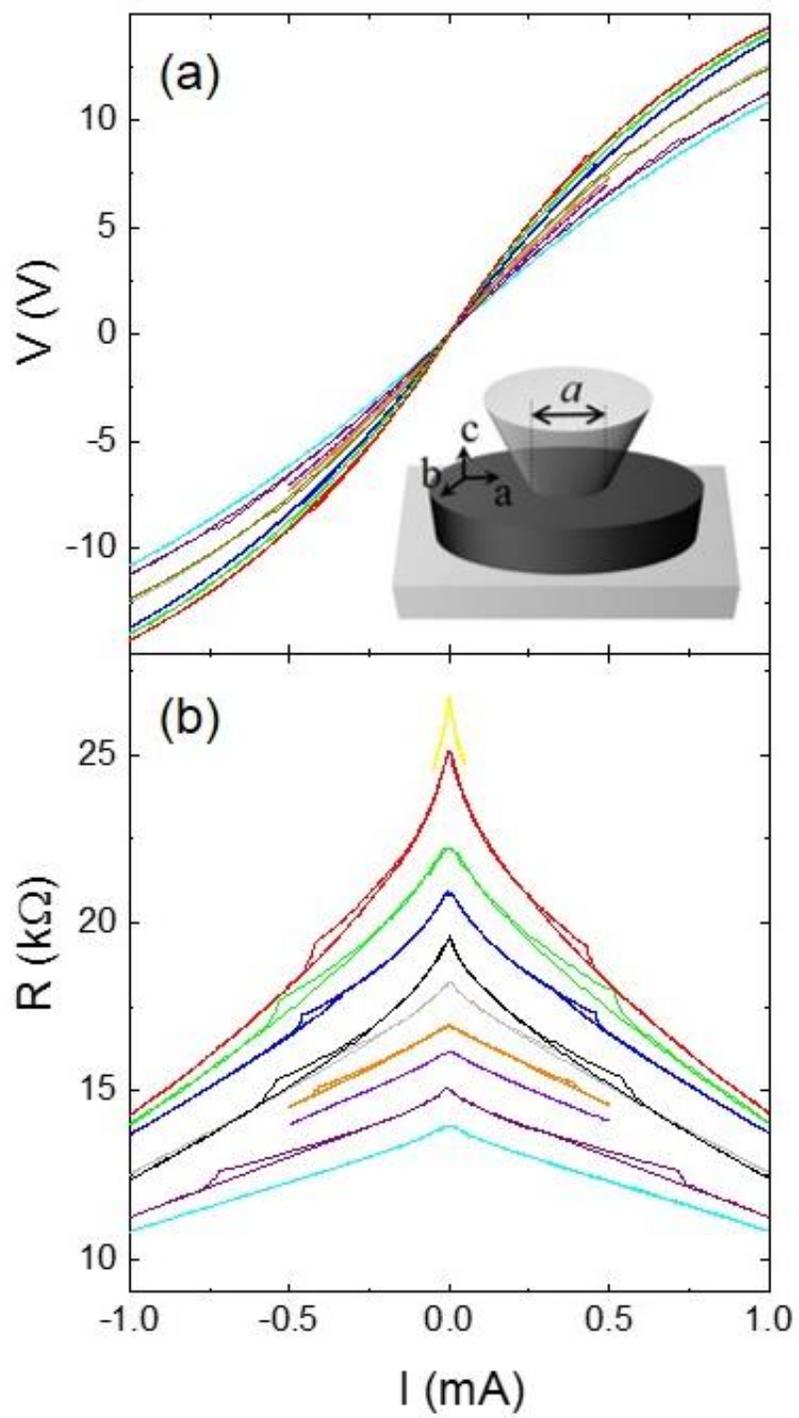

**Figure 2**

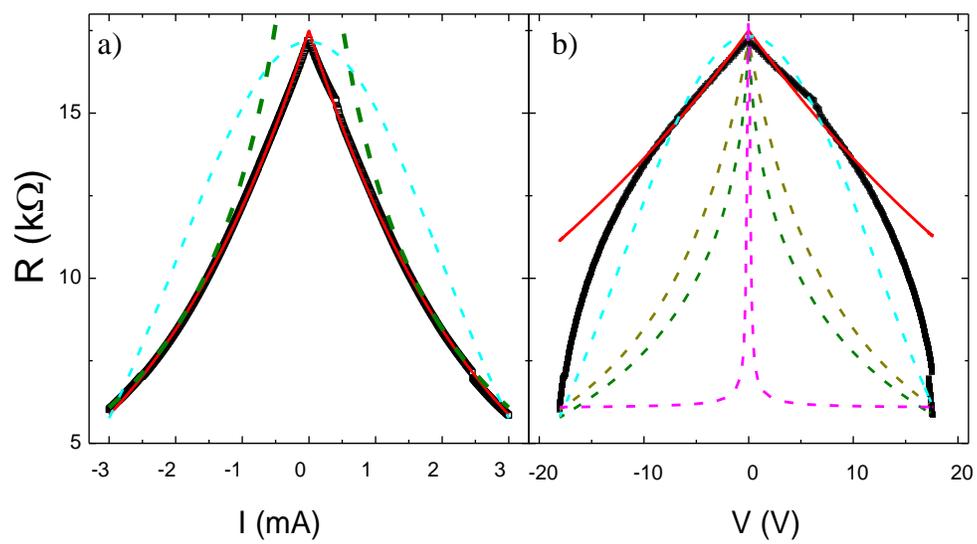

**Figure 3**

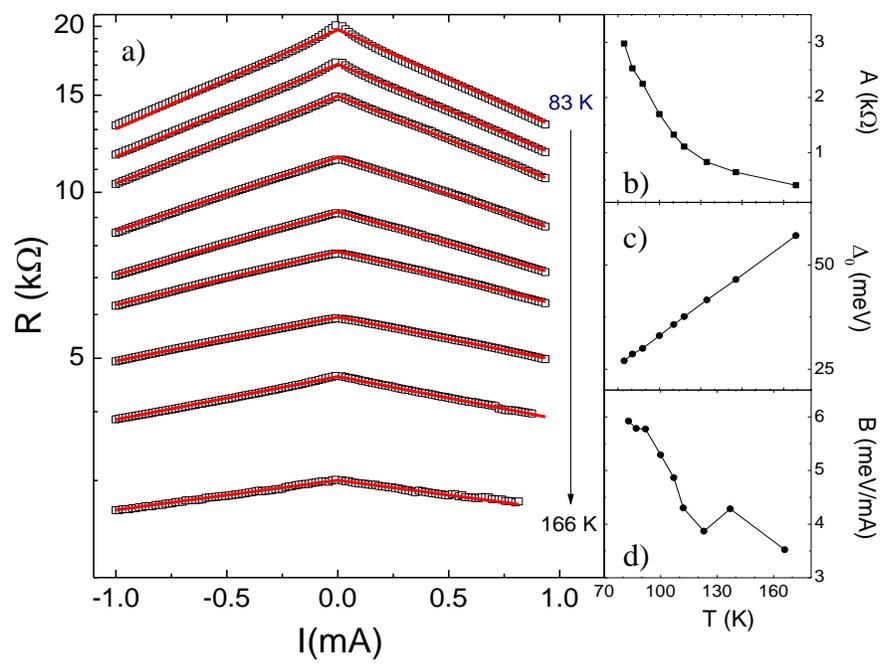

**Figure 4**

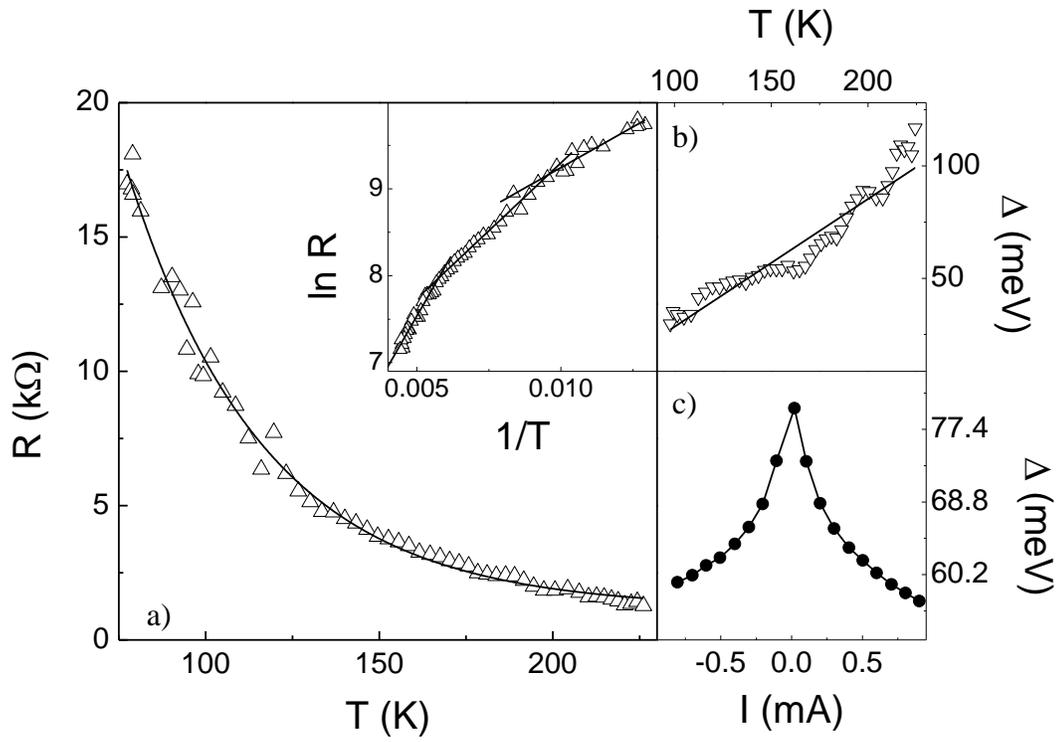

**Figure 5**

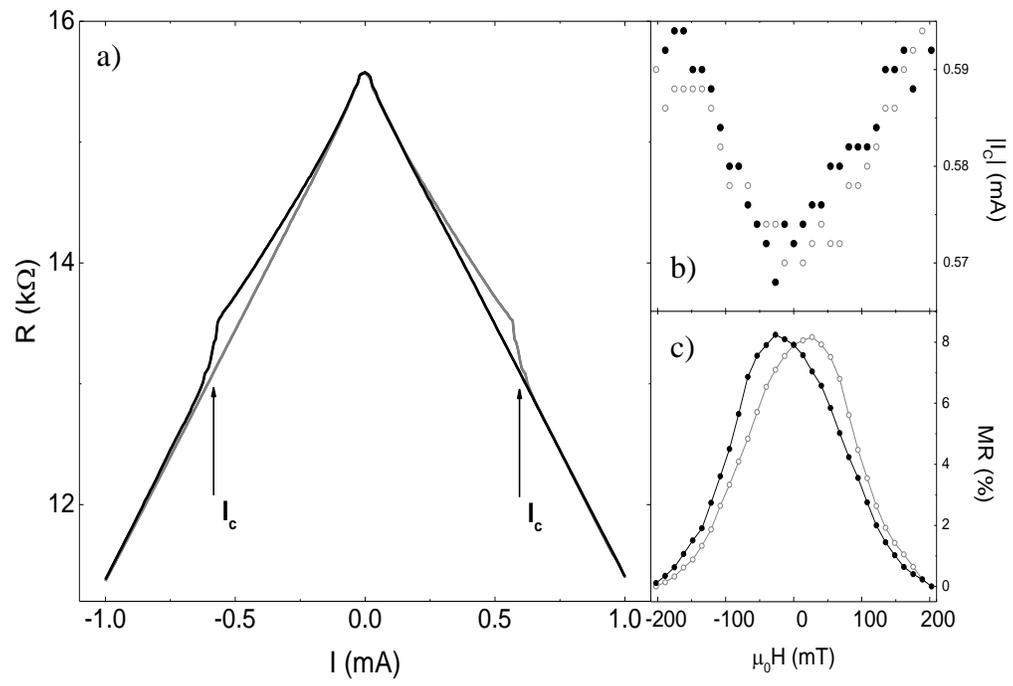

**Figure 6**

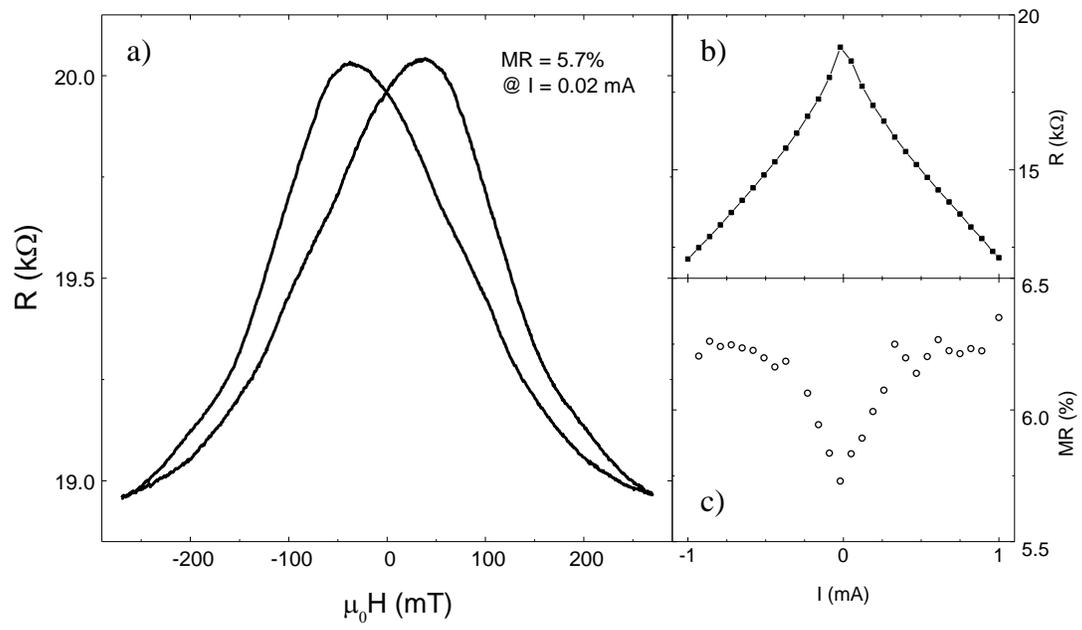